\documentclass[pre,aps,twocolumn,showpacs,amsmath,amssymb,amsfonts,floatfix,superscriptaddress]{revtex4}
\usepackage{graphicx}
\usepackage{dcolumn}
\usepackage{bm}
\usepackage[hypertex]{hyperref}
\usepackage{latexsym}
\usepackage{float}
\usepackage{supertabular}
\usepackage{longtable}

\begin{document}
\title{Stochastic dynamics of the prisoner's dilemma with cooperation facilitators}
\author{Mauro Mobilia}
\affiliation{Department of Applied Mathematics, School of Mathematics, University of Leeds, Leeds LS2 9JT, U.K.} %
\email{M.Mobilia@leeds.ac.uk}
\begin{abstract}
In the framework of the paradigmatic prisoner's dilemma game, we investigate the evolutionary dynamics of social dilemmas in
 the presence of ``cooperation facilitators''. In our model, cooperators and defectors interact as in the classic prisoner's 
dilemma, where selection favors defection. However, here the presence of a small number of cooperation facilitators enhances the fitness (reproductive potential) of cooperators, while it does not alter that of defectors.
In a finite population  of size $N$, the  dynamics of the prisoner's dilemma with facilitators is 
characterized by the probability that cooperation takes over (fixation probability) and by the mean times to reach 
the absorbing states.
These quantities are computed exactly and using Fokker-Planck equations. Our findings, corroborated by stochastic simulations, 
demonstrate that the influence of facilitators crucially depends on the difference between their density $z$ and the 
game's cost-to-benefit ratio $r$. 
When  $z>r$, the fixation of cooperators is likely in a large population and, under  weak selection pressure, invasion and replacement
of defection by cooperation is  favored by selection 
if $b(z-r)(1-z)>N^{-1}$, where $0<b\leq 1$ is the cooperation payoff benefit. 
When $z<r$, the fixation 
probability of cooperators is exponentially enhanced by the presence of  facilitators but
defection is   the  dominating strategy.
\end{abstract}

\maketitle

\section{Introduction}
Understanding the origin of cooperative behavior is a central issue in the life and behavioral sciences, and has recently been 
listed among the major scientific puzzles to be elucidated~\cite{Pennisi}.
Evolutionary game theory (EGT) provides the ideal framework to study
the competition between species and there is a long tradition of modeling the evolution of cooperation using evolutionary games~\cite{EGT,Axelrod}.
In recent years, these processes have increasingly been investigated using the methods of statistical physics, see e.g.~\cite{EGT}
and references therein.
In EGT, successful species spread at the expense of the others, and each individual's reproductive potential (fitness) 
varies with the  population's composition that  continuously changes in time. The interaction between the species is thus accounted for 
by a fitness-dependent (or ``frequency-dependent'') selection pressure~\cite{EGT}, as observed in various experiments~\cite{freqsel}. Quite intriguingly,
in such a setting the optimization of the fitness at an individual level can result in the reduction of the population
overall fitness~\cite{EGT,Axelrod}. An influential example of such a  paradoxical behavior,  
is provided by the celebrated  {\it prisoner's dilemma} (PD) game that serves as
 a metaphor for social dilemmas. In fact, in the classic PD individual interest leads to defection, 
even though mutual cooperation would be socially more beneficial~\cite{Axelrod,EGT}.
While the PD is the paradigmatic model for the evolution of cooperation, its main 
prediction is at odds with the cooperative behavior that is commonly observed
in experimental realizations~\cite{PDexp,freqsel}. This has motivated an upsurge of research 
 aiming to identify the possible
mechanisms capable of promoting cooperation in biological and social systems~\cite{coop}.
Notably, it has  been proposed that cooperation can be promoted by 
kin  and  group selection~\cite{group},
as well as by conditional behavioral rules leading to direct or indirect reciprocity~\cite{reciprocity,indirect_rec}.
 It has also been found that local interactions may promote cooperation in some social dilemmas~\cite{local}.
Furthermore, it has been shown that cooperation is supported in games with voluntary participation, or
with punishment for non-cooperation~\cite{other}.

In this work, we investigate an alternative scenario for the spread of cooperation in social dilemmas: we
consider the evolution of the prisoner's dilemma in a finite population comprising a  small number 
of ``cooperation facilitators''.
The facilitators  participate in the dynamics only by enhancing the reproductive potential of cooperators, while they do 
not affect the fitness of defectors (see Sec.~II below).
To study the influence of  cooperation facilitators on the prisoner's dilemma dynamics, the evolution is modeled in terms 
of a birth-death process and the fixation properties are studied 
analytically. 
In fact, it is well established that the evolutionary dynamics in finite populations is 
efficiently characterized by the probabilities of reaching the absorbing states, where the extinction of one 
or more species and the fixation of another occur~\cite{EGT,Kimura,weak,Antal06,weaksel}. 
 Here, we are  particularly interested in the probability that,  from a given initial 
composition, the population eventually   comprises only
 cooperators and a small fraction of facilitators,  but no defectors (``cooperation fixation probability'').
The mean times for these events (mean fixation times) are also studied and our results are checked against 
stochastic simulations. This approach allows us to (i) discuss how demographic fluctuations alter the mean field predictions 
of the classic replicator 
equations~\cite{EGT},  and (ii) thoroughly analyze the circumstances under which facilitators and
selection  favor a single cooperator
invading and replacing  a population of defectors.

This paper is organized as follows: The PD  with cooperation facilitators is introduced in the next section, 
where some of its properties are discussed. In Section III the  dynamics with the Fermi process
is characterized by
the fixation probability (Sec.~III.A) and the mean fixation times (Sec.~III.B). The dynamics
with the Moran process is studied in Section IV, while we summarize 
our findings and present our conclusions in Section V.
\section{Prisoner's dilemma with cooperation facilitators: model and dynamics}
In  evolutionary game theory, two-player games can be interpreted as dilemmas of cooperation.
In fact, the two possible strategies can be interpreted as ``cooperation'' (\textsf{C})
and ``defection'' (\textsf{D}). The paradigm of social dilemma is provided by the classic {\it prisoner's dilemma} (PD), 
whose main features are captured by the following payoff matrix giving the pairwise interaction between cooperators and 
defectors~\cite{EGT,Axelrod,local}\cite{switching}:
\begin{eqnarray}
 \label{payoff}
\bordermatrix{
  & \textsf{C} & \textsf{D} \cr
\textsf{C} & b-c & -c \cr
\textsf{D} & b & 0 \cr},
\end{eqnarray} 
where $b$ and $c$ respectively represent the benefit and the cost of cooperation, with $b>c>0$. 
Here, without loss of generality, we assume that $0<b\leq 1$.
According to (\ref{payoff}), mutual cooperation leads to a payoff $b-c>0$ and mutual defection gives a payoff $0$;
whereas when one player defects and the other cooperates, the defector receives a payoff $b$ and the cooperators gets $-c$. 
In the (classic) PD, the dilemma arises from the fact that each individual is better off not cooperating, even 
though mutual cooperation enhances the population overall payoff. Hence, while cooperation is socially beneficial,
 defection is the only (strict) Nash equilibrium in the PD~\cite{EGT,Axelrod}.

In this work, we consider a {\it finite} population comprising $N$ individuals on a complete graph (no spatial structure).
The number of cooperators and defectors is respectively denoted by $j$ and $k$. In addition to 
 cooperators and defectors, we  consider
that the population also comprises a fixed (small) number $\ell$ of ``cooperation facilitators'' ($\ell \ll N$).
These facilitators  cooperate with $\textsf{C}-$players and therefore 
{\it enhance} the  reproductive potential  (fitness) of cooperators, while  they leave the fitness of defectors 
 unaltered, see below.
Hence, while the number of cooperators and defectors in the population changes in time ($j$ and $k$ vary),
the total number of cooperators and defectors $j+k=N-\ell$ is conserved. According to the tenets of EGT, 
the variation in time of the  number of 
cooperators and defectors  depends on their average payoffs, $\pi_{\textsf{C}}$ and $\pi_{\textsf{D}}$ respectively,
 obtained from the payoff matrix (\ref{payoff}). Here, since facilitators enhance $\pi_{\textsf{C}}$
by cooperating with $\textsf{C}$ individuals and have no (direct) influence on $\pi_{\textsf{D}}$, one has
\begin{eqnarray}
 \label{payoffs}
\pi_{\textsf{C}}&=& (b-c)\frac{j+\ell-1}{N-1}-c \frac{k}{N-1}\nonumber\\
\pi_{\textsf{D}}&=& b \frac{j}{N-1},
\end{eqnarray} 
where  we  have excluded self-interactions from the definition of the payoffs~\cite{EGT}.
The population average payoff is  given by $\bar{\pi}=(j\pi_{\textsf{C}}+
k\pi_{\textsf{D}})/N$.  
It is worth noticing that
the expression of $\pi_{\textsf{C}}$ now comprises a  term $(b-c)\ell/(N-1)>0$
reflecting the positive contribution of facilitators to the cooperators payoff.
In evolutionary dynamics, it is customary to add a baseline constant, here set to $1$, to 
the payoffs $\pi_{\textsf{C}/\textsf{D}}$ of the spreading species~\cite{EGT,weaksel},
yielding the fitness of species $\textsf{C}$ and $\textsf{D}$, respectively given by
\begin{eqnarray}
 \label{fitness}
f_{\textsf{C}}&=& 1+ \pi_{\textsf{C}}= 1+ b\left[\frac{j+\ell-r(N-1)-1}{N-1}\right]\nonumber\\
f_{\textsf{D}}&=& 1+ \pi_{\textsf{D}}=1+b \frac{j}{N-1},
\end{eqnarray} 
where, we have introduced the {\it cost-to-benefit ratio} $r\equiv c/b$ (with $0<r<1$)
and have used $k=N-j-\ell$. Similarly, the 
  average fitness of the entire population reads $\bar{f}= (jf_{\textsf{C}}+ kf_{\textsf{D}})/N=1+[b(1-r)j - \ell]/N$ and 
grows linearly 
with the density $x\equiv j/N$ of cooperators.

The size of the population being finite, the evolutionary dynamics is modeled
as a continuous-time birth-death process~\cite{EGT,Gardiner,vanKampen}. 
In this model, only pairs of cooperators and defectors interact (according to (\ref{payoff}))
and the  stochastic dynamics is implemented as follows: (i) at each time step 
 a pair of individuals is randomly  chosen from the entire population; 
(ii) unless a pair of cooperator-defector is drawn, nothing happens; and
(iii) if one picks a cooperator-defector pair, one of these individuals
is randomly chosen for reproduction (proportionally to its fitness) and the other is replaced by the
newborn offspring. Hence, at each interaction the number of cooperators increases or decreases by one. 
The time evolution of this birth-death process can therefore be 
described by the random variable $j$ giving the number of cooperators 
and by the rates $T_j^{\pm}$ associated with the transitions $j \to j\pm 1$, respectively.
Here, we  consider 
\begin{eqnarray}
 \label{rates}
T_j^{\pm}=\frac{j(N-\ell-j)}{N(N-1)}\, \Psi^{\pm}(f_{\textsf{C}},f_{\textsf{D}}),
\end{eqnarray}
where $j(N-\ell-j)/N(N-1)$ accounts for the 
probability of picking a cooperator-defector pair, while  $\Psi^{\pm}$ are functions of the fitnesses (\ref{fitness})
that encode the interactions (selection) according to the chosen ``microscopic'' update rule~\cite{EGT}. 
We  here discuss the cases where $\Psi^{\pm}$ correspond to (i) the Fermi
process (FP)~\cite{FermiProcess,MA10}  and (ii) the Moran process (MP)~\cite{Moran,Kimura,EGT,Antal06,MA10} that are commonly used 
in EGT~\cite{EGT}.

Stochastic evolutionary dynamics and the influence of selection are generally characterized by the fixation properties,
 namely the probability that 
a given species fixates (takes over) the whole population and by the mean time for such an event to occur~\cite{EGT,Kimura,weaksel}.
In the absence of facilitators,  fixation happens when only one species survives and 
the population composition is uniform. Here, as the number of facilitators remains constant,
fixation will be achieved when one of the absorbing states is reached and
 either all cooperators are replaced by defectors, or vice versa, resulting in a (non-uniform) population 
comprising $\ell$ facilitators and $N-\ell$ cooperators or defectors.
In this work, we are particularly interested in the probability $\phi_j^{\textsf{C}}$
that, starting with $j$ cooperators, all defectors are eventually removed from the population and replaced by cooperators.
As discussed in Sec.~III.A., the fixation probability $\phi_j^{\textsf{C}}$ is necessary to establish
when selection favors cooperation replacing defection~\cite{weaksel}. 
In the framework of the above birth-death process (\ref{rates}), this probability
 obeys the  backward master equation~\cite{vanKampen,Antal06,EGT}
\begin{eqnarray}
 \label{backME}
\phi_j^{\textsf{C}}=T_j^-\phi_{j-1}^{\textsf{C}}+T_j^+\phi_{j+1}^{\textsf{C}}+[1-T_j^- - T_j^+]\phi_j^{\textsf{C}}, 
\end{eqnarray} 
with absorbing boundaries $\phi_0^{\textsf{C}}=0$ and $\phi_{N-\ell}^{\textsf{C}}=1$.
The formal solution of Eq.~(\ref{backME}) reads~\cite{vanKampen,Antal06,EGT}
\begin{eqnarray}
 \label{formal}
\phi_j^{\textsf{C}}=\frac{1+\sum_{n=1}^{j-1}\prod_{m=1}^{n}{\left(T_j^-/T_j^+\right)}}
{1+\sum_{n=1}^{N-\ell-1}\prod_{m=1}^{n}{\left(T_j^-/T_j^+\right)}}.
\end{eqnarray} 
Since the above birth-death process is a one-dimensional Markov chain, other quantities like the mean fixation times (MFTs)
can, in principle, be obtained exactly, but yield unwieldy expressions~\cite{vanKampen,Antal06}. 
When the population size $N$ is large, it is often much more useful to describe the fixation properties in terms of the diffusion approximation 
obtained in the continuum limit ($N\gg 1$) by a second-order size-expansion of the master equation resulting in a Fokker-Planck 
equation~\cite{Gardiner,vanKampen,Kimura,weak}. 
By denoting $x\equiv j/N$ and $y\equiv k/N$ the initial density of cooperators and defectors, respectively; 
and with $z\equiv \ell/N$ being the fraction of facilitators in the population, 
the (backward) Fokker-Planck equation (FPE) associated with (\ref{backME}) reads~\cite{Gardiner,vanKampen}
\begin{eqnarray}
 \label{G}
{\cal G}_{{\rm back}}(x)\phi^{\textsf{C}}(x)=0,
\end{eqnarray} 
where $\phi^{\textsf{C}}(x)\equiv \phi^{\textsf{C}}_{j/N}$ and
\begin{eqnarray}
 \label{bFPE}
{\cal G}_{{\rm back}}(x)&\equiv& [T^+(x)-T^-(x)]\frac{d}{d x} \nonumber\\ &+&
\frac{1}{2N} [T^+(x)+T^-(x)]\frac{d^2}{d x^2},
\end{eqnarray} 
with $T^{\pm}(x)\equiv T_{j/N}^{\pm}$ and, as usual,  
  the density
$x$ changes by $\pm \delta=\pm N^{-1}$ at each  cooperator-defector interaction.
In the realm of the Fokker-Planck equation, the timescale is such that the time-step is $\delta=N^{-1}$.
The formulation in terms of the FPE  allows a neat connection with the mean field treatment of the dynamics.
In fact, when $N\to \infty$ and all demographic fluctuations are negligible, the time variation of the density of cooperators 
is given by
the drift term of (\ref{bFPE})~\cite{Gardiner}, i.e.
\begin{eqnarray}
 \label{RE}
\frac{d x(t)}{dt}&=&T^+(x)-T^-(x)\\ &=&
x(1-z-x)[\Psi^+(f_{\textsf{C}},f_{\textsf{D}}) - \Psi^-(f_{\textsf{C}},f_{\textsf{D}})].\nonumber
\end{eqnarray} 
As for the classic PD, this rate equation admits two absorbing fixed points, $x=0$ (no cooperators) 
and $x =1-z$ (no defectors),
but possesses no interior fixed point since
$\Psi^+(f_{\textsf{C}}(x),f_{\textsf{D}}(x)) \neq \Psi^-(f_{\textsf{C}}(x),f_{\textsf{D}}(x))$  
for the Fermi and Moran processes, see below. As discussed in what follows,  the stability of these fixed points depends on the difference between the cost-to-benefit 
ratio $r$ and the fraction $z$ of facilitators.
\section{Dynamics with the Fermi Process}
The stochastic dynamics of evolutionary games is often conveniently modeled in terms of the so-called Fermi process (FP), see, for example,~\cite{FermiProcess}. In the FP, at each time-step two individuals are randomly drawn from the entire population and
one of them reproduces at the expense of the other that is replaced by the newborn offspring. This happens with a probability
proportional to the difference between the fitness of the interacting individuals and given by the Fermi function from statistical physics.
Since only the $\textsf{C}\textsf{D}$ pairs interact, the dynamics with the FP is described by 
the birth-death process defined by (\ref{rates}) and $\Psi^{\pm}=[1+e^{\mp(f_{\textsf{C}}-f_{\textsf{D}})}]^{-1}=
[1+e^{\mp(\pi_{\textsf{C}}-\pi_{\textsf{D}})}]^{-1}$~\cite{FermiProcess}. With these expressions of $\Psi^{\pm}$, 
one checks that
$\Psi^{+}(f_{\textsf{C}},f_{\textsf{D}})\neq \Psi^{-}(f_{\textsf{C}},f_{\textsf{D}})$  
(since $f_{\textsf{C}}\neq f_{\textsf{D}}$, see Eqs.~(\ref{fitness}),(\ref{fitMoran})), which confirms the absence of an interior fixed point
in the mean field (continuum) limit.
\subsection{Fixation probability}
With (\ref{fitness}), the transition rates (\ref{rates}) for the Fermi process read
\begin{eqnarray}
 \label{ratesFP}
T_j^{\pm}=\frac{j(N-\ell-j)}{N(N-1)}\,\frac{1}{1+{\rm exp}{\left(\pm v_N\right)}},
\end{eqnarray} 
with
\begin{eqnarray}
 \label{vN}
v_N\equiv f_{\textsf{D}}-f_{\textsf{C}}=b\left[r-\left(z-\frac{1}{N}\right)\left(1+\frac{1}{N-1}\right)\right]
\end{eqnarray} 
This quantity measures the selection pressure~\cite{special}.
 Clearly, $-bz<v_N<b(1-z)$ and $|v_N|\leq 1$.
In the continuum limit  $N\gg 1$, the densities $x=j/N, z=\ell/N$, and $v_N\to v\equiv b(r-z)$ are  treated as continuous quantities,
and the absence of self-interaction is ignored yielding the transition rates (\ref{ratesFP})
\begin{eqnarray}
 \label{ratesFPcont}
T^{\pm}(x)=\frac{x(1-z-x)}{1+e^{\pm v}}. 
\end{eqnarray} 
The  rate equation corresponding to the mean field dynamics with the Fermi process is obtained by using (\ref{ratesFPcont})
into (\ref{RE}), and is characterized by a single stable (absorbing) fixed point corresponding to a stationary density
\begin{eqnarray}
\label{fixpt}
 x^*= \left\{
  \begin{array}{l l}
    x_{\textsf{C}}=1-z  & \; \text{(no defectors)} \quad  \text{if $v<0$}\\
    x_{\textsf{D}}=0   & \;  \text{(no cooperators)} \quad  \text{if $v>0$}\\
  \end{array} \right.
\end{eqnarray}
of cooperators. This means that cooperation prevails ($x^*=x_{\textsf{C}}$, {\it no defectors}) in an infinitely large population 
comprising a fraction $z$ of facilitators 
 higher than  the cost-to-benefit ratio $r$, i.e. when $v<0$. However, as in the traditional PD,
defection wins ($x^*=x_{\textsf{D}}$, {\it no cooperators})  if $z$  is less than $r$  ($v>0$).
In other words, for cooperation to prevail ($x^*=x_{\textsf{C}}$) at mean field level, it is necessary that the density of facilitators  
 compensates the cost of cooperation relative to its benefit. 

When the population size is {\it finite}, demographic fluctuations are important and the evolution  is  thus  no longer aptly  
described by the mean field dynamics (\ref{RE}).
 In particular, the mean field results (\ref{fixpt})
do not account for the nonzero probability that a single 
cooperator can invade and replace a (finite) population of defectors.
Here, to investigate how the above mean field picture (\ref{RE}),(\ref{fixpt}) is altered by fluctuations arising in a finite population, we compute the 
probability $\phi_j^{\textsf{C}}$ that defection is eventually replaced by cooperation
in a population comprising  initially $j$  cooperators and $N-\ell -j$ defectors. 
Since  $T_j^{+}/T_j^{-}=e^{v_N}$,
this probability can be obtained  explicitly using (\ref{formal}) and, when 
$v_N\neq 0$~\cite{special}, one finds
\begin{figure}
\includegraphics[width=3.6in, height=2.4in,clip=]{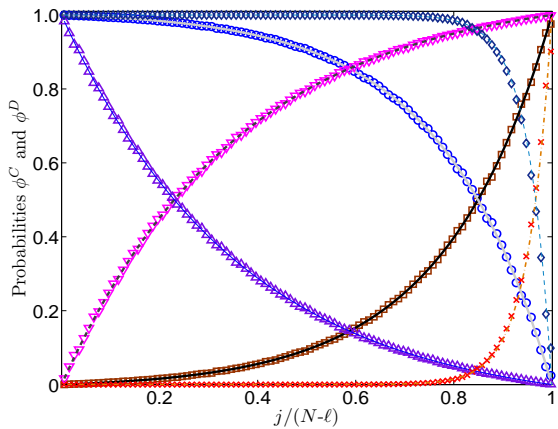}
\caption{{\it (Color online)}. Probabilities $\phi_j^{\textsf{C}}$ and 
$\phi_j^{\textsf{D}}$  for various  $z=\ell/N$
as function of $j/(N-\ell)=x/(1-z)$, and evolution with
the Fermi process. Results of stochastic simulations (symbols) for 
$\phi_j^{\textsf{C}}$ are compared with the predictions (curves) of (\ref{FixFP})
for $z= 0~(\times,$ dot-dashed),~$0.08~(\square$, solid),~$0.12~(\nabla$, dashed).
Similarly for $\phi_j^{\textsf{D}}$ with  $z=0~(\diamond,$ thin dashed),~$0.08~(\circ$, solid gray),~$0.12~(\triangle,$ thin solid).
The other parameters are $N=200, b=1.0, c=0.1$ (i.e. $r=0.1$).
 Stochastic simulations are for the birth-death process defined by (\ref{ratesFP}) and
 have been averaged over $2\times 10^5$ samples.
 }
 \label{PDzel_fig1}
\end{figure}
\begin{eqnarray}
 \label{FixFP}
\phi_j^{\textsf{C}}=\frac{e^{jv_N}-1}{e^{N(1-z)v_N}-1},
\end{eqnarray} 
while fixation probability of defectors  is simply given 
by  $\phi_j^{\textsf{D}}\equiv 1-\phi_j^{\textsf{C}}=(e^{N(1-z)v_N}-e^{jv_N})/(e^{N(1-z)v_N}-1)$.
As shown in Fig.~\ref{PDzel_fig1},   where results
 of stochastic simulations  
obtained using  the Gillespie algorithm~\cite{Gillespie}
 are reported, the predictions of (\ref{FixFP}) are in excellent agreement with numerical 
simulations. The expression (\ref{FixFP})  implies that 
\begin{eqnarray}
\label{FixFPas}
\phi_j^{\textsf{C}} \simeq \left\{
  \begin{array}{l l}
1-e^{-j|v_N|} & \quad \text{if $v_N<0$}\\
   (e^{jv_N}-1)~e^{-N(1-z)v_N} & \quad \text{if $v_N>0$},\\
  \end{array} \right.
\end{eqnarray}
when $N|v_N|\gg 1$.
In particular, the cooperation fixation probability starting with a single cooperator reads
\begin{eqnarray}
 \label{Fix1FP}
\phi_1^{\textsf{C}}&=&\frac{e^{v_N}-1}{e^{N(1-z)v_N}-1}\nonumber\\
&\simeq& \left\{
  \begin{array}{l l}
1-e^{-|v_N|} & \quad \text{if $v_N<0$}\\
    (e^{v_N}-1)~e^{-N(1-z)v_N}  & \quad \text{if $v_N>0$}.\\
  \end{array} \right.
\end{eqnarray} 
The findings (\ref{FixFP})-(\ref{Fix1FP}), summarized in  Fig.~\ref{PDzel_fig1}, illustrate how a small fraction $z$ of 
cooperation facilitators affects the fixation probabilities in a large, yet finite, 
population with an initial density of cooperators comparable to, or larger than, the density of 
  defectors: When $v_N<0$ ($z>r$),
the fixation probability of cooperators is much higher than that of defectors, 
$\phi^{\textsf{C}}_j\gg \phi^{\textsf{D}}_j$, and the spread of cooperation
is thus efficiently promoted by facilitators. Yet, it is worth noticing that 
 defectors still have finite probability to fixate even when $z>r$  (and $x\ll 1-z$), contrary to the mean field prediction (\ref{fixpt}).
The opposite situation arises when $v_N>0$ ($z<r$), as shown in Fig.~\ref{PDzel_fig1}.

\begin{figure}
\includegraphics[width=3.6in, height=2.4in,clip=]{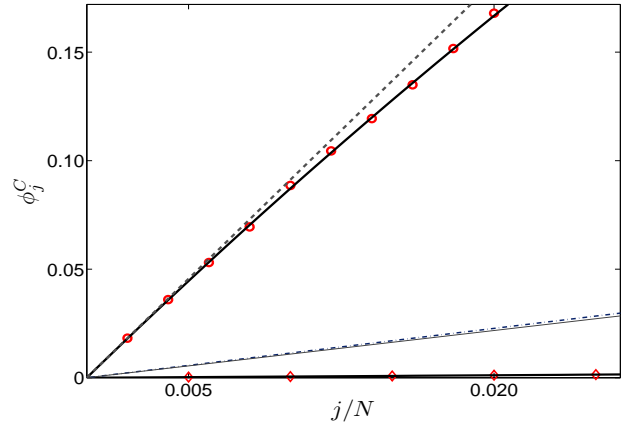}
\caption{{\it (Color online)}. Probability $\phi_j^{\textsf{C}}$ as function of 
$j/N$
when the initial number of cooperators is $j=1-10$ with $N=500$, and  $j=1-5$ with $N=200$. The dynamics is implemented according 
to the Fermi Process with  (\ref{ratesFP}).
The results of stochastic simulations (symbols, averaged over $2\times 10^5$ samples) 
are compared with (\ref{FixFP}) (curves/lines).  
Parameters are $b=1.0, c=0.1$ (i.e. $r=0.1$), and 
 $(N,z)=(500,0.12)~(\circ)$, $(200,0.08)~(\diamond)$. Here, $\phi_1^{\textsf{C}}\simeq 0.0182~(\circ)$ 
 and $\phi_1^{\textsf{C}}\simeq 2.88\times 10^{-4}~(\diamond)$, see text.
The dashed/dotted/thin lines correspond to  $\phi_j^C\simeq j v_N/(e^{N(1-z)v_N}-1)$ with $(N,z)=(500,0.12)$ (dashed) and 
 $\phi_{j,v_N=0}^C=j/(N-\ell)$ for $(N,\ell)=(200,16)$ (thin) and $(N,\ell)=(500,60)$ (dashed-dotted).
 }
 \label{PDzel_fig1smallx}
\end{figure}

The results (\ref{FixFPas}) and (\ref{Fix1FP})
can also be used to assess the influence of selection on the evolutionary dynamics~\cite{EGT}:
Following the seminal work of Ref.~\cite{weaksel}, we can establish when
selection favors cooperation (${\textsf{C}}$) {\it invading and replacing} defection (${\textsf{D}}$). 
Selection is  said to
favor the {\it replacement of ${\textsf{D}}$ by  ${\textsf{C}}$}
 if the fixation probability $\phi_1^{\textsf{C}}$ of a single cooperator in a
 population of $N-\ell-1$ defectors is greater than in absence of selection pressure ($v_N=0$) when $\phi_{1,v_N=0}^{\textsf{C}}=
(N-\ell)^{-1}$~\cite{special}. 
With (\ref{Fix1FP}), this yields the condition $1-e^{-|v_N|}>(N(1-z))^{-1}$ that
is generally satisfied  in  large populations under non-vanishing selection pressure.
An interesting result arises when the selection intensity is weak  and the population size is large, i.e.
$|v_N|\to |v|\ll 1$ and $N\gg 1$. In such a limit, $\phi_{1}^{\textsf{C}}\simeq |v|$ when $z>r$ 
(see Fig.~\ref{PDzel_fig1smallx} where $|v|=0.02$) and selection favors cooperation replacing
defection provided that $b(z-r)>(N(1-z))^{-1}$.
Moreover, selection favors ${\textsf{C}}$ {\it invading} ${\textsf{D}}$ 
when $f_{\textsf{C}}>f_{\textsf{D}}$~\cite{weaksel}. With 
(\ref{fitness}), this yields the condition $z-r>(1-r)/N$.
Therefore, under weak selection ($|v|\ll 1$ and $N\gg 1$) selection favors the invasion and replacement of ${\textsf{D}}$
by ${\textsf{C}}$ if
 $z-r>\frac{1}{N}\, {\rm max}\left(1-r, \frac{1}{b(1-z)}\right)$.
Since $0<b\leq 1$, one has $1-r\leq (b(1-z))^{-1}$ and cooperation invading and replacing defection is favored by 
selection provided that
\begin{eqnarray}
 \label{selfav}
b(z-r)(1-z)>N^{-1}.
\end{eqnarray} 

One can also use the results  (\ref{Fix1FP}) to determine the circumstances under which defection is {\it evolutionary stable}.
In fact according to \cite{weaksel}, and as natural extension of the concept of evolutionary stability for infinitely large populations 
and deterministic evolutionary dynamics~\cite{EGT}, ${\textsf{D}}$ is evolutionary stable in a {\it finite} population if 
(i) selection opposes ${\textsf{C}}$
 invading ${\textsf{D}}$, implying $f_{\textsf{C}}<f_{\textsf{D}}$, i.e.  $z-r<(1-r)/N$; {\it and} if
(ii) selection opposes ${\textsf{C}}$
 replacing ${\textsf{D}}$, i.e. $\phi_1^{\textsf{C}}<(N-\ell)^{-1}$.
The condition (ii) is clearly always satisfied when $|v_N|$ is finite, and in this case defection is evolutionary stable if 
$z-r<(1-r)/N$. In the weak selection limit where $|v|\ll 1$ (with $N\gg 1$), the condition (ii) yields $z-r<(bN(1-z))^{-1}$.
 Hence, defection is evolutionary stable  under weak selection in a large population if 
$z-r<\frac{1}{N}\, {\rm min}\left(1-r, \frac{1}{b(1-z)}\right)=(1-r)/N$. Since $r<1$, this clearly
 means that defection is evolutionary stable and is the dominating  strategy  when $z<r$.
It is worth noticing that in the limit of an infinite population, $N\to \infty$, one recovers the mean field results (\ref{fixpt}):
 cooperation prevails only if $z>r$, according to (\ref{selfav}), and defection dominates otherwise.

The meaning of the results (\ref{FixFPas})-(\ref{selfav}) is illustrated in Fig.~\ref{PDzel_fig1smallx}
where $\phi_{j}^{\textsf{C}}$ has been computed in populations comprising 
  a small initial number of cooperators ($j=1, ..., 10$) and an excellent agreement with (\ref{FixFPas}) and 
(\ref{Fix1FP}) has been found.  In Fig.~\ref{PDzel_fig1smallx}, $|v_N|\ll 1$ and  
we notice that 
 $\phi_{j}^{\textsf{C}}$ increases linearly in $x=j/N\ll 1$, 
with a slope steeper than $(1-z)^{-1}$ when  $z>r$
and selection favors cooperation replacing defection. 
The slope is less than $(1-z)^{-1}$ when  $z<r$ and the fixation of cooperation is opposed by selection.
To further appreciate the implications of (\ref{FixFP})-(\ref{selfav}), 
it is useful to compare these results with those obtained in  the absence of facilitators.
Putting $z=0$ in (\ref{FixFP})-(\ref{selfav}), one recovers the results for the classic PD
when cooperation fixation probability 
vanishes exponentially with the population size $N$:
 $\phi_{j,z=0}^{\textsf{C}}\sim e^{-(N-j)c}$ and 
$\phi_{1,z=0}^{\textsf{C}}\sim e^{-Nc}$~\cite{EGT,Antal06} (see Fig.~\ref{PDzel_fig1}).

Our findings therefore demonstrate that facilitators greatly influence the probability that cooperation prevails
and are summarized  in Fig.~\ref{PDzel_fig1}.
As illustrated in that figure, the influence of facilitators crucially depends on the difference between their density
 $z$ and the cost-to-benefit ratio $r$:
\begin{enumerate}
 \item[i.] When $v_N<0$, the fixation of cooperators is likely  (but not certain) even when they are initially in minority, i.e.
even when initially $x=j/N<1/2$. 
\item[ii.]  When $v_N<0$ and $N|v_N|\gg 1$, the fixation probability of a single cooperator is 
generally higher than in the absence of selection pressure  ($v_N=0$). In this case, with $z>r$, selection {\it favors}  cooperation invading and replacing 
defection, see (\ref{Fix1FP}) and 
Figs.~\ref{PDzel_fig1} and \ref{PDzel_fig1smallx}. Furthermore, under weak selection pressure and in a large 
population ($|v|\ll 1$ and $N\gg 1$), 
 the fixation probability of a single cooperator is 
independent of $N$, $\phi_{1}^{\textsf{C}}\simeq z-r$. In this case invasion and replacement of defection by cooperation is 
favored by selection if (\ref{selfav}) is satisfied.
\item[iii.] When $v_N>0$, selection always opposes cooperation replacing defection. In this case,  
while defection is evolutionary stable and is the dominating strategy when $z<r$,
the cooperation fixation probability is  exponentially enhanced by a small fraction of facilitators. Yet,  
cooperation is likely to fixate only if defectors are initially outnumbered 
by cooperators, i.e. if $j\gg k$, as illustrated in Fig.~\ref{PDzel_fig1}.
\end{enumerate}
\subsection{Mean fixation times}
Another quantity of great interest to unveil the influence
of facilitators in the evolutionary dynamics of the PD
is the (unconditional) {\it mean fixation time}. This quantity gives the average time necessary to reach one of 
the absorbing boundaries, i.e. a population composition with either $0$ or $N-\ell$ cooperators.
The unconditional mean fixation time (MFT), $\tau_j$, for a system comprising initially $j$ cooperators 
obeys the following backward master equation~\cite{vanKampen,Antal06,EGT} (where the time-step is $\delta=N^{-1}$)
\begin{eqnarray}
 \label{backMFT}
\tau_j=\delta+ T_j^-\tau_{j-1}+T_j^+\tau_{j+1}+[1-T_j^--T_j^+]\tau_j,
\end{eqnarray} 
with boundary conditions $\tau_0=\tau_{N-\ell}=0$. In principle, this equation can be solved exactly but the final result is cumbersome and not enlightening. Here, in the continuum limit  $N\gg 1$, we work with the continuous 
quantities $x=j/N, z=\ell/N$ and $v= b(r-z)$, and adopt the approach of the {\it diffusion theory}~\cite{Kimura,Gardiner}.
The diffusion approximation is known to be particularly suited to analyze the dynamics under weak selection, which here corresponds 
to the regime where $|v|\ll 1$~\cite{weaksel,EGT,weak}. Exact methods (when available) or other approximations~\cite{Antal06}, {\it e.g.} 
the WKB approach~\cite{MA10}, are particularly useful to deal with the case of strong selection intensity and/or with phenomena like metastability.
 In the realm of the diffusion theory, the transition rates of the 
FP are given by (\ref{ratesFPcont})
and the fixation probability of cooperation is obtained by solving  (\ref{bFPE}) which yields
\begin{eqnarray*}
\phi^{\textsf{C}}(x)=\frac{e^{Nvx}-1}{e^{N(1-z)v}-1},
\end{eqnarray*}
while for defection the  probability is
$\phi^{\textsf{D}}(x)=1-\phi^{\textsf{C}}(x)$.

\begin{figure}
\includegraphics[width=3.6in, height=2.4in,clip=]{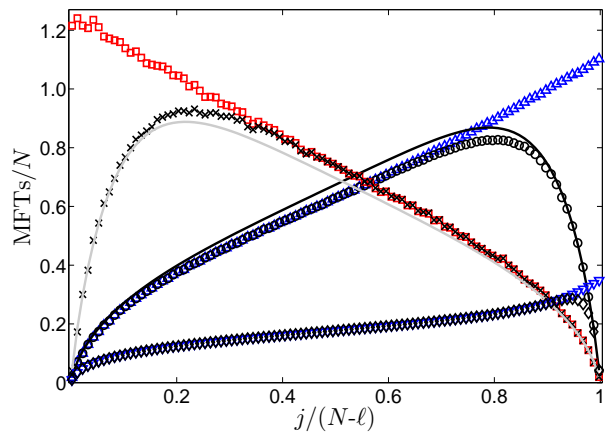}
\caption{{\it (Color online)}.
 Mean fixation times as function of $j/(N-\ell)=x/(1-z)$ for the evolution with  the Fermi process. 
Results of stochastic simulations (symbols) for $\tau$ are compared with the solution (curves) of 
Eq.~(\ref{bFPE_FP}) for $z=0~(\diamond),~0.08~(\circ$, solid black),~$0.12~(\times,$ solid gray). 
We also report the numerical results for the  conditional MFTs $\tau^{\textsf{C}}$ for 
$z=0.12~(\square)$ and  $\tau^{\textsf{D}}$ with $z=\ell/N=0~(\nabla),~0.08~(\triangle)$. The other parameters are $N=500, b=1.0, c=0.1$, i.e. $r=0.1$. Stochastic simulations are for the FP  with rates (\ref{ratesFP}) and have 
been averaged over $2\times 10^5$ samples.
 }
 \label{PDzel_fig2}
\end{figure}

Similarly, the unconditional MFT is obtained by solving the backward FPE ${\cal G}_{{\rm back}}(x) \tau(x)=-1$~\cite{Gardiner,weak}, i.e.
\begin{eqnarray}
 \label{bFPE_FP}
x(1-z-x)\left[\tanh{\left(\frac{v}{2}\right)} \,\frac{d}{dx} - \frac{1}{2N} \frac{d^2}{dx^2} \right]\tau(x)=1,
\end{eqnarray} 
with the absorbing boundary conditions $\tau(0)=\tau(1-z)=0$. When the  drift and  diffusive terms are of the same order, 
 i.e. when $|v|\sim N^{-1}\ll 1$, it follows from Eq.~(\ref{bFPE_FP}) that the MFT scales linearly with $N$, i.e.
\begin{eqnarray}
\label{scaling}
 \tau(x)=N{\cal F}_v(x).
\end{eqnarray}
The scaling function can be obtained explicitly by 
solving Eq.~(\ref{bFPE_FP}) using standard methods, see e.g.~~\cite{Gardiner}.
For instance, when the initial density of cooperators and defectors is the same, $x=y=(1-z)/2$, and $|v|\sim N^{-1}\ll 1$, one finds
\begin{eqnarray}
\label{Fv}
{\cal F}_v\left(\frac{1-z}{2}\right)&=&\frac{e^{-(2+z)q}}{q(1-z)(1+e^{-q(1-z)})}\left[e^{(2+z)q} \right.\nonumber\\
&\times& \left. \left\{\gamma_{{\rm E}} - \ln{2} -{\rm Ei}(-(1-z)q)\right\}
+e^{(1+2z)q} \right. \nonumber\\ 
&\times& \left.
\left\{{\rm Ei}((1-z)q)
 - \gamma_{{\rm E}} - \ln{(2q)} \right\} \right.
\\
&+& \left. e^{3zq}~\left\{
{\rm Ei}((1-z)q)-{\rm Ei}(2(1-z)q)
\right\}
 \right. \nonumber\\
&+& \left. e^{3q}~\left\{
{\rm Ei}(-2(1-z)q)-{\rm Ei}(-(1-z)q)
\right\}
\right. \nonumber\\ 
&+& \left.
 e^{(1+z)q}~\left(e^q -e^{zq}\right)~\ln{(1-z)}
 \right],  \nonumber
\end{eqnarray}
where $q\equiv N|\tanh{(v/2)}| \simeq N|v|/2$,  ${\rm Ei}(x)\equiv \int_{-\infty}^x\frac{e^u}{u}\,du$ 
denotes the usual exponential integral and $\gamma_{{\rm E}}=0.5772...$ is Euler-Mascheroni constant.
While the expression of ${\cal F}_v$ is usually cumbersome, some useful properties can be directly inferred 
from (\ref{bFPE_FP}). In fact, as Eq.~(\ref{bFPE_FP}) is invariant under 
the transformation $(x,r)\to (1-z-x, 2z-r)$, one has
 ${\cal F}_v(x)={\cal F}_{-v}(1-z-x)$ when $z$ is kept fixed. 
The unconditional MFT in the Fermi process is  therefore characterized by the symmetry
\begin{eqnarray}
\label{symmetry1}
 \tau(x)|_{r}=\tau(1-z-x)|_{r \to r'=2z-r},
\end{eqnarray}
where, on the right-hand-side 
$r$ is replaced by $r'=2z-r$ and $v$ transformed into $-v$, with $z$ kept fixed.
Furthermore, when $r=c/b$ is kept fixed but $z$ varies, (\ref{bFPE_FP}) is
invariant under the transformation $z\to z'=2r-z$ and $x\to 1-z'-x$, while the boundary conditions become
$\tau(1-z')=0$ and $\tau(z-z')=\tau(-2v/b)=0$. In the weak selection regime $|v|/b=|z-r|\ll 1$, 
the second boundary condition can be approximated by
$\tau(z-z')\simeq\tau(0)=0$, which allows a mapping onto (\ref{bFPE_FP}) that
yields:
\begin{eqnarray}
\label{symmetry}
 \tau(x)|_{z}\simeq \tau(1-z'-x)|_{z\to z'=2r-z},
\end{eqnarray}
 with $r=c/b$ fixed. The comparison between the solution of (\ref{bFPE_FP}) and the results of stochastic simulations 
(for the FP with rates (\ref{ratesFP}))
reported in Fig.~\ref{PDzel_fig2}  shows that the diffusion approximation aptly captures the 
functional dependence of $\tau$, even though some deviations (of about $10\%$) can be noticed. 
These deviations stem from the self-interaction terms that are excluded from (\ref{ratesFP}) but not in the continuum limit
 (\ref{ratesFPcont}) [e.g. in Fig.~\ref{PDzel_fig2} one has $v_N\simeq -0.0182$ and $v=-0.02$ when $z=0.12$, and
$v_N\simeq 0.0218$ and $v=0.02$ for $z=0.08$]. More importantly, the scaling (\ref{scaling}) and 
the relationship (\ref{symmetry}) are confirmed by the numerical simulations of Fig.~\ref{PDzel_fig2}.
In fact,  in Fig.~\ref{PDzel_fig2} we notice that $\tau(x)$ is a humped function  with a maximum well separated from the absorbing boundaries and
located at  $x/(1-z)<1/2$ when $z>r$ and, while $\tau$ scales linearly with $N$, the presence of facilitators increases 
the unconditional MFT and its maximum value at the hump.

In addition to the unconditional MFT, it is also relevant to consider the mean time to specifically 
reach one of the absorbing boundaries.
Hence, the  conditional mean fixation times $\tau^{\textsf{C}}(x)$ and 
$\tau^{\textsf{D}}(x)$ respectively 
give the average time  
to  reach the absorbing boundaries
$x=1-z$ and $x=0$~\cite{Antal06,MA10}. As for the unconditional MFT,  
these quantities can be obtained from a backward FPE in the realm of the diffusion approximation. In fact, $\tau^{\textsf{C}}(x)$
obeys  ${\cal G}_{{\rm back}}(x) [\phi^{\textsf{C}}(x)\tau^{\textsf{C}}(x)]=-\phi^{\textsf{C}}(x)$, with 
the absorbing boundaries $\phi^{\textsf{C}}(1-z)\tau^{\textsf{C}}(1-z)=\phi^{\textsf{C}}(0)
\tau^{\textsf{C}}(0)=0$~\cite{Kimura}.
Since $\phi^{\textsf{D}}(x)=1-\phi^{\textsf{C}}(x)$
and,  from (\ref{FixFP}), $\phi^{\textsf{D}}(x)=\phi^{\textsf{C}}(1-(2r-z)-x)$, 
the conditional MFTs  in the regime $|v|/b\ll 1$ (weak selection)
are related  by the relationship $\tau^{\textsf{C}}(1+z-2r-x)|_{z}\simeq \tau^{\textsf{D}}(x)|_{z \to z'= 2r-z}$ where $r$ is kept fixed, as illustrated in Fig.~\ref{PDzel_fig2}.
Furthermore,  one has $\phi^{\textsf{D}}(x)\simeq 1$ when $x\to 0$ and $v>0$ ($z<r$),
 while  $\phi^{\textsf{C}}(x)\simeq 1$ when $x\to 1$ and $v<0$ ($z>r$). This implies that
$\tau(x) \simeq 
\left\{
  \begin{array}{l l}
\tau^{\textsf{D}}(x) & \quad \text{when $v>0$ and $x\to 0$}\\
 \tau^{\textsf{C}}(x) & \quad \text{when $v<0$ and $x\to 1$}.\\
  \end{array} \right.$
 As shown 
in Fig.~\ref{PDzel_fig2}, 
  $\tau^{\textsf{C}}(x)$ decreases while  $\tau^{\textsf{D}}(x)$ increases
monotonically with $x/(1-z)$.

The influence of facilitators on the unconditional and conditional MFTs is summarized in Fig.~\ref{PDzel_fig2}.
We have found that in the PD with cooperation facilitators  all conditional and 
unconditional MFTs scale linearly with the population size $N$ when $|v|\sim N^{-1}$ (weak selection). While  a similar scaling is also obtained in the absence of facilitators, the 
MFTs at a fixed value $x/(1-z)$ are found to be significantly increased by
 the presence of facilitators.
Hence, the  presence of cooperation facilitators has the quantitative effect  to prolong the coexistence and the competition between  
cooperators and defectors before an absorbing state is reached, see Fig.~\ref{PDzel_fig2}.
\section{Dynamics with the fitness-dependent Moran process}
The stochastic dynamics of evolutionary games is often implemented in terms of the Moran process, see 
e.g.~\cite{weaksel,EGT},
that was originally introduced in  population genetics \cite{Moran,Kimura}. In its essence, the Moran model is 
a birth-death process where one randomly picked individual produces an offspring proportionally to its fitness
relative to the population average fitness.
The resulting offspring then replaces another individual that is randomly picked to be removed from the population
 whose size is therefore
conserved. Here, as the interactions are between cooperators and defectors, the Moran process is implemented with 
$\Psi^+=f_{\textsf{C}}/\bar{f}$ and  $\Psi^-=f_{\textsf{D}}/\bar{f}$
in (\ref{rates}). Since $f_{\textsf{C}}\neq f_{\textsf{D}}$ when $v\neq 0$ [see (\ref{fitMoran}) and \cite{special}], 
one verifies that 
$\Psi^{+}(f_{\textsf{C}},f_{\textsf{D}})\neq \Psi^{-}(f_{\textsf{C}},f_{\textsf{D}})$  implying
 the absence of an interior fixed point in the mean field (continuum) limit.
The Moran process is usually investigated when the selection intensity is weak, both for technical convenience 
(the mathematical treatment simplifies greatly) and for the biological relevance of such a limit~\cite{EGT,Kimura,weaksel}. 
In this section, the stochastic dynamics with the Moran process is investigated 
in the weak selection limit, where  $|v|=b|r-z|\ll 1$,  using the diffusion approximation.
\subsection{Fixation probability}
In the continuum limit, the fitnesses (\ref{fitness}) become 
\begin{eqnarray}
\label{fitMoran}
 f_{\textsf{C}}(x)=1-v+bx \quad \text{and} \quad  f_{\textsf{D}}(x)=1+bx,
\end{eqnarray}
with $\bar{f}(x)=1-z+b(1-r)x$.
The transition rates for the Moran process thus read
\begin{eqnarray}
\label{rates_Moran}
T^{+/-}(x)=x(1-z-x)\frac{f_{\textsf{C}/\textsf{D}}(x)}{\bar{f}(x)}. 
\end{eqnarray}
With (\ref{fitMoran}) and (\ref{rates_Moran}), the mean field dynamics is described by the rate equation (\ref{RE})
whose properties are similar to those discussed for the Fermi process. In particular, the rate equation (\ref{RE}) for 
the Moran process is also characterized by a single stable (absorbing) fixed point $x^*=x_{\textsf{C}}$ (no defectors) if $v<0$ and 
$x^*=x_{\textsf{D}}$ (no cooperators)  if $v>0$ [see (\ref{fixpt})].

To understand how the combined effect of nonlinear selection and demographic noise alters the mean field description,
we now compute the cooperation fixation probability in the realm of the diffusion approximation.
In such a setting, the fixation probability $\phi^{\textsf{C}}(x)$
is given by the FPE (\ref{G},\ref{bFPE}) with the boundary conditions $\phi^{\textsf{C}}(0)=0$ and $\phi^{\textsf{C}}(1-z)=1$.
The solution of  (\ref{G}) is given by ~\cite{Gardiner}
\begin{eqnarray}
\label{Phi_Moran}
\phi^{\textsf{C}}(x)=\frac{\int_0^x du \; \chi(u)}{\int_0^{1-z} du  \; \chi(u)},
\end{eqnarray}
where, with (\ref{fitMoran}) and (\ref{rates_Moran}),
\begin{eqnarray}
\label{chi}
 \chi(u)&=&{\rm exp}\left(-2N \int_0^u ds \left\{\frac{T^{+}(s)-T^{-}(s)}{T^{+}(s)+T^{-}(s)}\right\}\right)\nonumber\\
&=& {\rm exp}\left(2Nv \int_0^u \; \frac{ds}{2bs+2-v} \right).
\end{eqnarray}
\begin{figure}
\includegraphics[width=3.6in, height=2.4in,clip=]{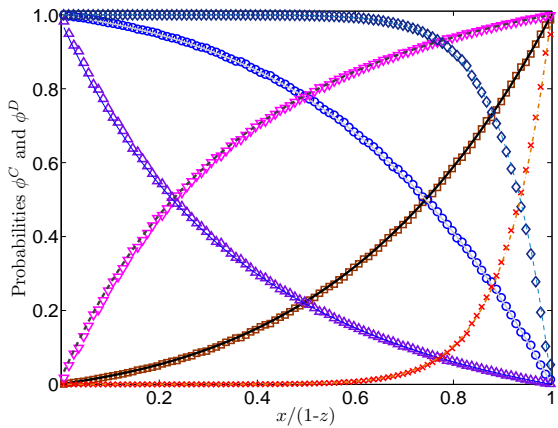}
\caption{{\it (Color online)}. Probabilities $\phi^{\textsf{C}}(x)$ and 
$\phi^{\textsf{D}}(x)$ for various $z$
as functions of $x/(1-z)$, and the dynamics with the Moran process (\ref{rates_Moran}). Results of 
stochastic simulations (averaged over $2\times 10^5$ samples)
are compared with (\ref{Phi_M}) 
for $z=0~(\times,$ dot-dashed),~$0.08~(\square$, solid), $0.12~(\nabla$, dashed).
Similarly for $\phi^{\textsf{D}}(x)$ with $z=0~(\diamond,$ thin dashed),~$0.08~(\circ$, solid gray),  $0.12~(\triangle,$ thin solid).
The other parameters are $N=200, b=1.0, c=0.1$ ($|v|=0.02$).
 }
 \label{PDzel_fig3}
\end{figure}
Introducing (\ref{chi}) into (\ref{Phi_Moran}) and performing the integrals, one obtains
\begin{eqnarray}
\label{Phi_M}
\phi^{\textsf{C}}(x)=\frac{\left(1+\frac{2b}{2-v}\,x\right)^{1+Nv/b} - 1}{\left(1+\frac{2b}{2-v}\, (1-z)\right)^{1+Nv/b}-1}.
\end{eqnarray}
As shown in Fig.~\ref{PDzel_fig3}, this
 result is in excellent agreement with numerical simulations
and exhibits the same qualitative features obtained for the Fermi process 
(compare with Fig.~\ref{PDzel_fig1}). The finding (\ref{Phi_M}) 
 implies that in the weak selection limit where $|v|\ll 1$ and  $N|v|\gg 1$, one has
\begin{eqnarray}
\label{Phi_M_as}
\phi^{\textsf{C}}(x) \simeq \left\{
  \begin{array}{l l}
1-(1+bx)^{-N(z-r)} & \quad \text{if $z>r$}\\
    \left(\frac{1+b(x-\frac{z-r}{2})}{1+b(1-\frac{z+r}{2})}\right)^{N(r-z)} & \quad \text{if $r>z$}.\\
  \end{array} \right.
\end{eqnarray}
In particular, the probability that cooperation fixates starting with a single cooperator, 
when $z>r$ 
is given by $\lim_{Nx\to 1}\phi^{\textsf{C}}(x)= 1-e^{-|v|}\simeq |v|$. 
We therefore recover the result derived from (\ref{Fix1FP}) for the Fermi process. 
Clearly, this implies that under weak selection the fixation of a single cooperator is favored by selection  if 
the non-trivial condition (\ref{selfav}) is satisfied. 
Again, it is instructive to compare  (\ref{Phi_M}), (\ref{Phi_M_as}) with the result obtained in 
the absence of facilitators, 
when $\phi^{\textsf{C}}(x)|_{z=0}\simeq \left(\frac{1+b(x-r/2)}{1+b(1-r/2)}\right)^{Nr}$
 decays to zero exponentially with $N$. The  influence of the facilitators on the fixation probabilities
for the Moran process is summarized in Fig.~\ref{PDzel_fig3}, where  the  same features as in
  Fig.~\ref{PDzel_fig1} are recognized and  summarized as follows: 
\begin{enumerate}
 \item[i.]  The fixation of cooperators is likely (but not certain) when the density of facilitators is higher than the cost-to-benefit ratio ($z>r$).
\item[ii.] When $|v|\ll 1$ and $N|v|\gg 1$, selection favors cooperation invading and replacing defection  
if  (\ref{selfav}) is satisfied. In particular, 
the fixation probability of a single cooperator is  $\lim_{Nx\to 1}\phi^{\textsf{C}}(x)\simeq |v|$.
\item[iii.] When $z<r$, selection opposes cooperation replacing defection 
but the fixation probability of cooperators is exponentially enhanced 
by the presence of facilitators.%
\end{enumerate}
\subsection{Mean fixation times}
In the realm of the diffusion approximation, the unconditional mean fixation time $\tau$
obeys the backward FPE
${\cal G}_{{\rm back}}(x) \tau(x)=-1$, 
with the absorbing boundary conditions $\tau(0)=\tau(1-z)=0$.
In the weak selection regime $c<b\ll 1$ and 
continuum limit,
 with (\ref{rates_Moran}), one has
\begin{eqnarray*}
\label{drift-diff}
T^+(x) -T^-(x)
&\simeq& -\frac{v}{1-z}\,x(1-z-x),\nonumber\\
T^+(x) +T^-(x) 
&\simeq& \frac{2}{1-z}\,x(1-z-x).
\end{eqnarray*}
With these expression, the backward FPE for the unconditional MFT 
reads
\begin{eqnarray}
 \label{bFPE_Moran}
\frac{x(1-z-x)}{1-z}\left[-v \,\frac{d}{dx} + \frac{1}{N} \frac{d^2}{dx^2} \right]\tau(x)=-1,
\end{eqnarray} 
with $\tau(0)=\tau(1-z)=0$. When $|v|, b\ll 1$, Eq.~(\ref{bFPE_Moran})
coincides with the FPE (\ref{bFPE_FP}) for the Fermi process 
with an effective 
population size $N(1-z)/2$. The solution to (\ref{bFPE_Moran}) can therefore readily be obtained from
(\ref{bFPE_FP}) and (\ref{scaling}). In particular, we infer from
 (\ref{scaling}) that 
the MFT scales linearly with $N(1-z)/2$
when $|v|\sim N^{-1}$, 
yielding
\begin{eqnarray}
\label{MFT_Moran_FP}
 \tau(x)=\frac{N(1-z)}{2}{\cal F}_{v}(x),
\end{eqnarray}
 where ${\cal F}_{v}(x)$ is the scaling function  
(\ref{scaling}) obtained for the 
 Fermi process.
This function   still satisfies the symmetry
 ${\cal F}_{v}(x)={\cal F}_{-v}(1-z-x)$ yielding
$ \tau(x)|_{r}=\tau(1-z-x)|_{r \to r'=2z-r}$, when $z$ is kept fixed, as in the Fermi process.
\begin{figure}
\includegraphics[width=3.6in, height=2.4in,clip=]{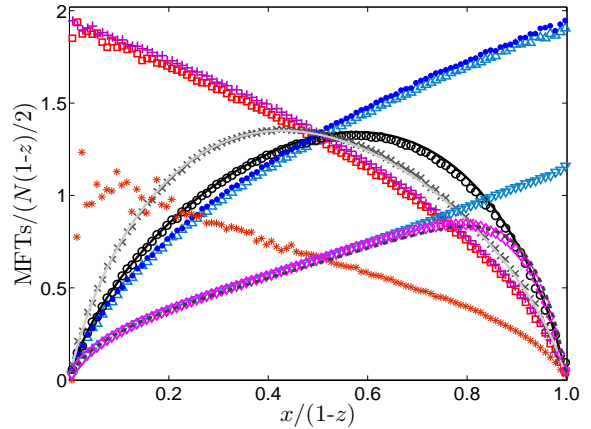}
\caption{{\it (Color online)}. Rescaled mean fixation times  as function of $x/(1-z)$ 
for the evolution with the Moran process~(\ref{ratesFPcont}). Results of stochastic simulations 
 for 
$\tau$ are compared with~(\ref{bFPE_Moran})
for $z=0~(\diamond$, dashed),~$0.16~(\circ$, solid black curve), $0.24~(\times,$ solid gray). 
Numerical results for the  conditional MFTs $\tau^{\textsf{C}}$ with $z=0~(*),~0.16~(\square), 0.24~(+)$
and for  $\tau^{\textsf{D}}$ 
with $z= 0~(\nabla),~0.16~(\triangle), 0.24~(\bullet)$.
The parameters are $N=200, b=0.25, c=0.05$ ($r=0.2$ and $|v|=0.04$). Stochastic simulations have been averaged over $2\times 10^5$ samples
 }
 \label{PDzel_fig4}
\end{figure}
In the same manner, from (\ref{MFT_Moran_FP}) and (\ref{symmetry}), we infer
\begin{eqnarray}
\label{symMoran}
 \tau(x)\simeq \left(\frac{1-z}{1+z-2r}\right)\tau(1-z'-x)|_{z\to z'= 2r-z}
\end{eqnarray}
when $r=c/b$ is kept fixed and $z$ is transformed into $z'= 2r-z$. 
The solution of (\ref{bFPE_Moran}), as well as the
 relationships (\ref{MFT_Moran_FP}) and (\ref{symMoran}), are in excellent agreement with the results of 
stochastic simulations reported
in Fig.~\ref{PDzel_fig4}. As for the FP, we can also consider the conditional mean fixation times
and it follows from (\ref{MFT_Moran_FP}) and (\ref{symmetry}) that  for the Moran process the conditional 
MFTs are related by
 $ \left(\frac{1+z-2r}{1-z}\right)\tau^{\textsf{C}}(1+z-2r-x)|_{z}\simeq \tau^{\textsf{D}}(x)|_{z \to z'= 2r-z}$ where $r$ is kept fixed,
in agreement with the results of  Fig.~\ref{PDzel_fig4}.
The influence of facilitators on the MFTs with the Moran process
is summarized in Fig.~\ref{PDzel_fig4}, where   the MFTs 
rescaled by a factor $(N(1-z)/2)^{-1}$ reproduce the same qualitative behavior obtained for the
Fermi process (compare with Fig.~\ref{PDzel_fig2}) and $\tau(x)$ is a humped function with a pronounced maximum. Again, all 
 MFTs scale linearly with  $N$ (in the weak selection 
limit). Yet, the comparison with the results for $z=0$ reveals that, at  a fixed value of $x/(1-z)$, the presence of facilitators  
increases the MFTs, see Fig.~\ref{PDzel_fig4}. Also, we notice that the monotonic dependence of 
  $\tau^{\textsf{C}}$  and  $\tau^{\textsf{D}}$ on $x$ 
is  essentially independent of the sign of $v\neq 0$ (in Fig.~\ref{PDzel_fig4}, $v=\pm 0.04$ and $v=0$).
\section{Summary and conclusion}
In this work, we have proposed and investigated  an alternative scenario leading to the spread of cooperation in social dilemmas.
We have considered the evolutionary dynamics of the  {\it prisoner's dilemma} (PD) game
in the presence of a small number of {\it cooperation facilitators}. These individuals participate in the 
dynamics only by enhancing the fitness of cooperators. The influence of  facilitators on the evolutionary dynamics has been
characterized by computing the model's fixation properties in a {\it finite} population of size $N$. Here, fixation 
occurs either in the state with only defectors (as in the classic PD),
or in the state where the entire population is comprised of cooperators and facilitators.
The dynamics has been implemented with the Fermi and Moran processes and the same qualitative results  have been found, 
which demonstrates the robustness of our findings. Our analytical  approach, corroborated by stochastic simulations, is based on an exact treatment and on the diffusion approximation (Fokker-Planck equation) of the underlying birth-death process.

Our main results concern the fixation probabilities, whose properties crucially depend on whether the fraction of
 facilitators $z$ is more or less than the game's cost-to-benefit ratio $r$. When $z>r$, 
we have shown that
facilitators are very efficient in promoting the spread of cooperators
 whose fixation is likely (but {\it not} certain, contrary to the mean field predictions) in a large population with 
comparable initial densities
of defectors and cooperators. Furthermore, when the selection intensity is weak and $N\gg 1$, we have 
demonstrated that the invasion and replacement of defectors by a {\it single cooperator}
 is favored by facilitators and selection  if $b(z-r)(1-z)>N^{-1}$ 
(where $0<b\leq 1$ is the cooperation payoff benefit). 
When $z<r$, defection is evolutionary stable and is the dominating strategy. In this case, while cooperation is unlikely to fixate, the 
fixation probability of cooperators is still exponentially enhanced by the presence of facilitators.  
 We have also studied the (unconditional and conditional)
 mean fixation times in the weak selection limit and found that  these quantities grow linearly with the population size. 
While  a similar scaling is also obtained in the absence of facilitators, their presence has the effect of significantly increasing 
all the  mean fixation times and hence to prolong the coexistence of cooperators and defectors.

In conclusion, this work demonstrates that the presence of a small number of cooperation facilitators can effectively enhance 
the spread of cooperation in a simple model of social dilemmas and  prolong the coexistence of competing species.
The influence of facilitators is particularly drastic when their abundance exceeds the game's cost-to-benefit ratio, in which case 
cooperation is generally the strategy favored by selection in large populations.
These findings pave the way to further investigations of 
the influence of facilitators  in other social dilemmas, e.g. with mixed strategies and/or in spatial settings. 


\begin{thebibliography}{99}
\bibitem{switching}
Such a formulation is commonly used 
since it represents the most  convenient way to capture 
 the essential features of  the classic PD ~\cite{local}. It is an example 
of ``equal gains from switching'' games, see e.g.~\cite{EGT} and references therein.
\bibitem{special}
When the density of facilitators is $z=r+(1-r)/N$, the selective pressure due to the cost-to-benefit ratio is balanced
 and one has $v_N=0$.
In this special case, the dynamics is neutral (there is effectively no selection) with 
$T^{+}_j/T^{-}_j=1$, and (\ref{formal}) yields 
the expected result $\phi_{j,v_N=0}^{\textsf{C}}=j/(N-\ell)=x/(1-z)$.
\end{thebibliography}
\end{document}